\documentclass[a4paper,11pt]{article}
\usepackage{pos}

\newcommand{\gmtwo}{\ensuremath{g\!-\!2}}
\newcommand{\amu}{$a_{\mu}$}

\title{Status of the Muon $g\!-\!2$/EDM Experiment at J-PARC}

\author[a,b,*]{Graziano Venanzoni}

\affiliation[a]{University of Liverpool, Liverpool L69 3BX, United Kingdom}

\affiliation[b]{INFN Sezione di Pisa, Largo Bruno Pontecorvo 3, 56127, Pisa, Italy}

\affiliation[*]{On behalf of the J-PARC $g\!-\!2$/EDM Collaboration}

\emailAdd{graziano.venanzoni@liverpool.ac.uk}

\abstract{
The Muon $g\!-\!2$/EDM Experiment at J-PARC will employ a novel way to measure the muon magnetic anomaly, $a_\mu = (g-2)_\mu/2$, by using a low-emittance beam of positive muons stored in a compact muon storage magnet. The experimental method includes new technologies such as a three-dimensional spiral injection, an MRI-type storage magnet with superb field uniformity, and a positron tracking detector. 
The expected systematic uncertainty will be at the same level as that of the Fermilab Muon $g\!-\!2$ experiment, providing an important cross-check of the “storage-ring method” employed at BNL and Fermilab. I will present the current status of the experiment, ongoing tests and design optimizations, and the plans for improvements of the experimental precision.
}

\FullConference{The European Physical Society Conference on High Energy Physics (EPS-HEP2025)\\
7-11 July 2025\\
Marseille, France\\}

\begin{document}
\maketitle

\section{Introduction}
The measurement of the muon magnetic anomaly $a_\mu = (g-2)_\mu/2$ provides one of the most precise tests of the Standard Model (SM)~\cite{Jegerlehner:2017gek,Aoyama:2020ynm}.  The long-standing tension between experiment and theory has driven major efforts on both sides, and the Fermilab Muon $g\!-2$ has measured $a_{\mu}$ to the impressive accuracy of 0.127 ppm~\cite{Muong-2:2025xyk}. 
In recent years, major progress in lattice-QCD determination of the leading-order hadronic vacuum polarization
reaching sub-percent accuracy~\cite{Borsanyi:2020mff}, has brought the Standard Model prediction much closer to the experimental value~\cite{Aliberti:2025beg}. At the same time, persistent tensions in the hadronic sector---both among different sets of $e^+e^-$ data 
used in the dispersive approach and between dispersive and lattice-QCD evaluations~\cite{Aliberti:2025beg}---currently limit the ultimate precision of the SM test. 
As has often been the case in the past, improved experimental measurements act as a catalyst for theoretical progress. 
Complementary efforts—spanning from lattice-QCD calculations to the refined analyses of present and forthcoming $e^+e^-$ cross-section data~\cite{Aliberti:2025beg,Aliberti:2024fpq}, as well as novel approaches such as the MUonE experiment at CERN~\cite{CarloniCalame:2015obs,Abbiendi:2016xup}—are expected to clarify the present situation.
In parallel, a new measurement of the muon \gmtwo\ employing a technique fundamentally different from the storage-ring method used at Brookhaven National Laboratory (BNL) and Fermilab is being prepared at J-PARC~\cite{Abe:2019thb}. The experiment will measure \amu\ (and the muon EDM) using an ultra-cold, reaccelerated muon beam stored in a compact 3-T magnetic ring, with a targeted systematic uncertainty competitive with the Fermilab Muon $g\!-\!2$ result.

\subsection{Experimental Approach}

The J-PARC muon $g\!-\!2$/EDM experiment (E34)~\cite{Abe:2019thb} is being developed at the Japan Proton Accelerator Research Complex (J-PARC), within the Materials and Life Science Experimental Facility (MLF). It makes use of the intense surface-muon beam produced at the MUSE H-line, providing a flux of approximately $1\times10^8$~$\mu^+$/s. The distinctive feature of this experiment is the production and reacceleration of a low-emittance muon beam, enabling a compact storage ring  and avoiding the presence of the electric field for vertical focusing as in the storage ring method~\cite{Gabrielse:2025jep}.

Surface muons with kinetic energy around 4~MeV are first stopped in a silica-aerogel target, where they form thermal muonium atoms ($\mu^+e^-$) in vacuum. The process converts a fraction of the surface muons into muonium with a nearly thermal momentum distribution. These muonium atoms are then ionized by a two-step laser excitation: the first laser with a wavelength of 122~nm (Lyman-$\alpha$) promotes the $1S \rightarrow 2P$ transition, and a second laser with a wavelength of 355~nm ionizes the excited state, liberating ultra-slow muons with kinetic energies of about 25~meV and 50\% polarization. This technique has been successfully demonstrated at J-PARC, providing a clean source of slow muons suitable for precision acceleration~\cite{Aritome:2024rlu}.

The ultra-slow muons are subsequently reaccelerated in a dedicated multi-stage linear accelerator system consisting of a radio-frequency quadrupole (RFQ), an inter-digital H-type drift tube linac (IH-DTL), and an Alvarez-type linac. The final beam momentum reaches approximately 300~MeV/$c$, while preserving the excellent emittance of the source. The resulting muon beam exhibits a transverse emittance reduced by about three orders of magnitude compared with conventional surface-muon beams, which is a crucial factor for achieving high injection efficiency and for minimizing betatron oscillations inside the storage region.

After reacceleration to 300~MeV/$c$, the beam with a 10 ns-wide pulses, consisting of three
microbunches, with a repetition rate of 25 Hz time structure,  is transported to the experimental area and injected into a compact superconducting solenoid magnet through a three-dimensional spiral injection scheme. This injection method allows the muons to adiabatically enter the storage region without the need for strong electric focusing fields, which are a major source of systematic uncertainties in previous storage-ring experiments. The storage magnet itself is based on an MRI-type superconducting solenoid with a diameter of about 66~cm, corresponding to roughly one-twentieth the size of the BNL and Fermilab rings. The field uniformity has been optimized to reach 1ppm 
across the storage volume, ensuring long spin-coherence times and a highly accurate determination of the spin-precession frequency.
For each pulse, or "fill", approximately about 100 muons are injected in the storage ring. Shortly after each fill, approximately 30 positrons from muon decays are detected, within a 5 ns time window,
by a cylindrical silicon-strip tracking detector surrounding the storage region. This detector provides high spatial and timing resolution, allowing precise reconstruction of the decay positron trajectories and time spectra. To ensure good
acceptance, the sensitive area along the axial direction is set to $\pm 200$ mm and the radial sensitive range is from
$90$ mm to $290$ mm. 
From the time-dependent distribution of detected positrons, both the anomalous spin-precession frequency (for $a_\mu$) and the up–down asymmetry (for EDM) are extracted. Dedicated readout electronics, including the custom SliT ASIC chip, have been developed to provide sub-nanosecond timing resolution and minimal time-walk effects, essential for precise event reconstruction at high counting rates.

\begin{figure}[htb]
    \includegraphics[width=.5\textwidth]{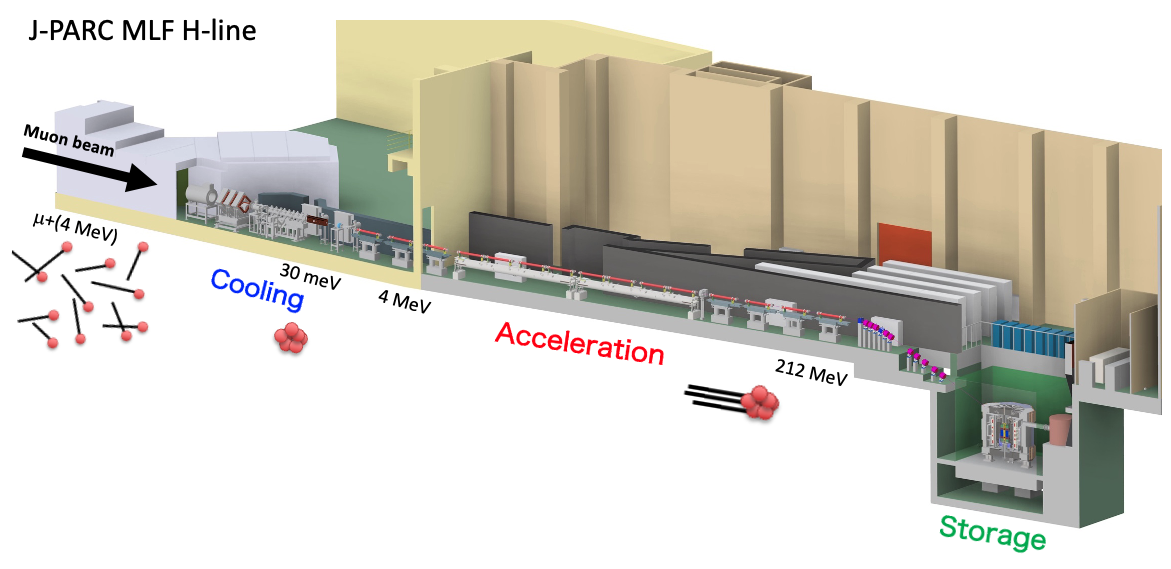}
    \includegraphics[width=.5\textwidth]{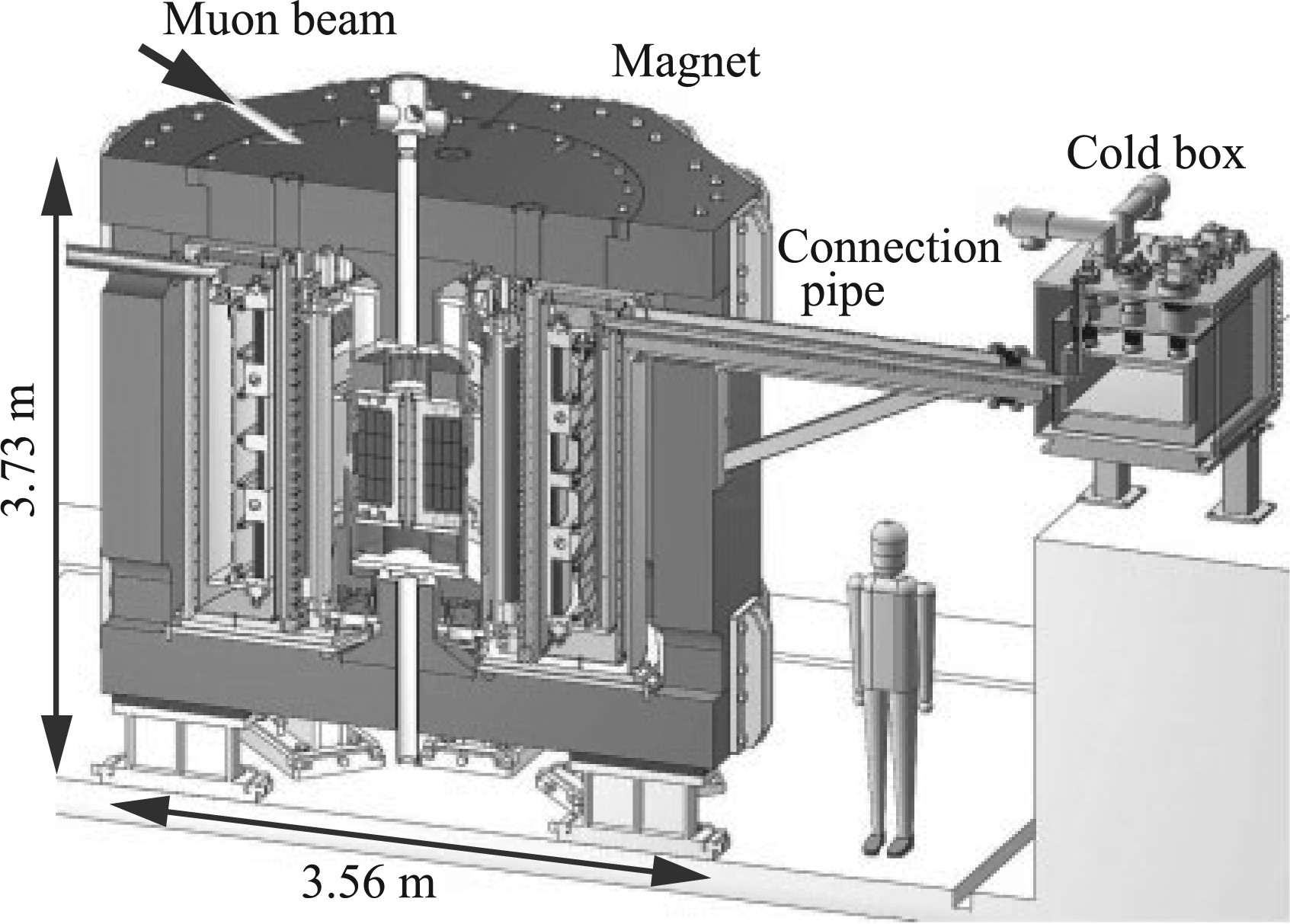}  
    \caption{Left: Schematic view of the accelerator complex for the muon $g-2$/EDM experiment at J-PARC (courtesy of T. Mibe). Right: Overview of the muon storage magnet (from~\cite{Abe:2019thb}).}
    \label{fig:jparc}
\end{figure}

Figure~\ref{fig:jparc} shows a schematic view of the accelerator complex and the muon storage magnet.
After two years of data taking $5.7 \times 10^{11}$ positrons are expected to be collected which should allow to reach a statistical accuracy of 450~ppb, while the systematic error will be kept to $<70$ ppb. The muon EDM goal is a statistical sensitivity of $1.5\times10^{-21}\ e\cdot\mathrm{cm}$  with a systematic uncertainty of $0.36\times10^{-21}\ e\cdot\mathrm{cm}$, a factor of 60
improvement over the present measurement~\cite{Muong-2:2008ebm}.

\subsection{Comparison of $g\!-\!2$ Experiments}

The experimental principle of muon $g-2$ measurements is based on determining the anomalous spin precession frequency 
$\vec{\omega}_a = \vec{\omega}_s - \vec{\omega}_c$, 
which, in the presence of electric and magnetic fields, can be expressed as
\[
\vec{\omega}_a = \frac{e}{m_\mu} 
\left[ a_\mu \vec{B} - \left(a_\mu - \frac{1}{\gamma^2 - 1}\right) \frac{\vec{\beta} \times \vec{E}}{c} \right].
\]
Here, $\vec{\omega}_s$ is the spin-precession frequency, $\vec{\omega}_c$ is the cyclotron frequency and $\vec{B}$ and $\vec{E}$ are the magnetic and electric fields, respectively. 

In the storage-ring experiments conducted at BNL and Fermilab, a strong electric field is used for vertical focusing and confinement of the muon beam. To suppress the contribution of the electric term in the equation above, the muons are stored at the so-called ``magic momentum'' ($p = 3.094$~GeV/$c$, $\gamma = 29.3$), where the factor $\left(a_\mu - \frac{1}{\gamma^2 - 1}\right)$ vanishes. The storage ring has a 14~m diameter with a magnetic field of about 1.45~T, and the experiment determines $a_\mu$ by correlating the time distribution of decay positrons above a certain energy threshold (like 1 GeV) with the precisely mapped magnetic field.

The J-PARC $g\!-\!2$/EDM experiment adopts a fundamentally different approach as discussed before.
In this setup, no electric field is used for focusing ($E = 0$), and vertical confinement is achieved magnetically (weak magnetic focusing with a field index $n\sim10^{-4}$). This configuration eliminates the need for the magic momentum condition and avoids the electric-field related systematic uncertainties present in storage-ring experiments.

Table~\ref{tabg-2} summarizes the main differences between the storage ring approach at BNL and Fermilab and the ultracold teachnique at J-PARC.

\begin{table}[h!]
\centering
\caption{Comparison of key parameters for the muon $g\!-\!2$ experiments at BNL (E821), Fermilab (E989), and J-PARC (E34).}
\label{tabg-2}
\begin{tabular}{lccc}
\hline\hline
 & BNL E821 & Fermilab E989 & J-PARC E34 \\
\hline
Muon momentum & $3.09$ GeV/$c$ & $3.09$ GeV/$c$ & $300$ MeV/$c$ \\
Lorentz factor $\gamma$ & $29.3$ & $29.3$ & $\sim 3$ \\
Muon lifetime in ring & $64.4~\mu$s & $64.4~\mu$s & $6.6~\mu$s \\
Beam polarization $P$ & $\sim 100\%$ & $97\%$ & $50\%$ \\
Typical asymmetry $A$ & $0.40$ & $0.40$ & $0.40$ \\
Storage magnetic field $B$ & $1.45$ T & $1.45$ T & $3.0$ T \\
Focusing method & Electrostatic & Electrostatic & Very weak magnetic \\
Ring diameter & $14$ m & $14$ m & $0.66$ m \\
Storage radius & 711 cm & 711 cm & 33.3 cm \\
Cyclotron period & $149$ ns & $149.1$ ns & $7.4$ ns \\
Spin precession period $T_{g-2}$ & $4.37~\mu$s & $4.40~\mu$s & $2.11~\mu$s \\
Spin precession frequency $\omega_a$ & $1.43$ MHz & $1.43$ MHz & $2.96$ MHz \\
Detected positrons $e^+$ & $8.5\times 10^{9}$ & $1.43\times 10^{11}$ & $5.7\times 10^{11}$ (proj.) \\[3pt]
Statistical goal (stat.) & $460$ ppb & $100$ ppb & $450$ ppb \\
Systematic goal (syst.) & $280$ ppb & $80$ ppb & $<70$ ppb \\
Events in final fit $N$ & $5\times10^{9}$ & $1.5\times10^{11}$ & $5.7\times10^{11}$ \\
EDM sensitivity (stat.) & $0.2\times 10^{-19}~e\cdot$cm & O($10^{-21})~e\cdot$cm & $1.5\times 10^{-21}~e\cdot$cm \\
EDM sensitivity (syst.) & $0.9\times 10^{-19}~e\cdot$cm & O($10^{-21})~e\cdot$cm  & $3.6\times 10^{-22}~e\cdot$cm \\
\hline\hline
\end{tabular}
\end{table}

A simple back-of-the-envelope estimate of the expected statistical precision can be derived from the relation:
\begin{equation}
    \frac{\delta\omega_a}{\omega_a} = 
    \frac{1}{\omega_a \gamma \tau_\mu P}
    \sqrt{\frac{2}{N A^2}},
    \label{eqn_sensi}
\end{equation}
where $\omega_a$ is the anomalous precession frequency, $\gamma\tau_\mu$ the dilated muon lifetime, $P$ the beam polarization, $A$ the decay asymmetry, and $N$ the total number of detected decay positrons used in the fit. Using the inputs from Table~\ref{tabg-2}, it gives the expected statistical sensitivity.

\section{Higher sensitivity studies for the J-PARC Muon $g\!-\!2$/EDM Experiment}

To further improve the statistical precision and reach of the muon $g\!-\!2$/EDM measurement at J-PARC, 
three main upgrade concepts are currently under investigation:
\begin{enumerate}
    \item Increasing the muon beam polarization through optical repolarization of muonium before ionization.
    \item Increase the beam momentum via a high-gradient muon linac up to 600 MeV/$c$.
    \item Strengthening the storage magnetic field beyond 3~T (possibly to 6~T).
\end{enumerate}

\begin{figure}[htb]
   \centering
    \includegraphics[width=.8\textwidth]{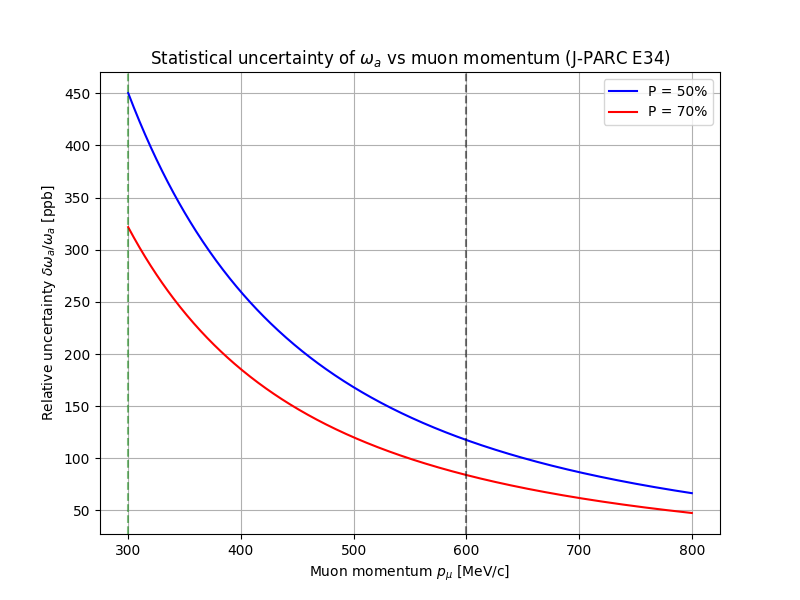}  
    \caption{Statistical uncertainty $\delta\omega_a/\omega_a$ vs.\ muon momentum for J-PARC E34, shown for beam polarizations of 50\% (blue) and 70\% (red). Dashed lines indicate a muon momentum of 300~MeV/c (current proposal) and a possible upgrade at 600~MeV/c. }
    \label{figsensi}
\end{figure}

The effect of each of these upgrades on the expected sensitivity can be easily understood by Equation~\ref{eqn_sensi}. Indeed, while the sensitivity scales linearly with the polarization $P$ (and with the square root of the number of events), it scales almost quadratically with the muon momentum\footnote{Assuming the muon orbit radius remains unchanged with respect to the 300 MeV/c design value.}, as shown in Fig.~\ref{figsensi}. Indeed assuming $5.7\times 10^{11}$ collected positrons, an increases of the momentum to 600 MeV/c (which requires a 6 T magnet) will bring the statistical sensitivity below 120 ppb, which can be further reduced to below 90 ppb by increasing the polarization to 70\%.

Each of these options addresses different aspects of the experimental system and has distinct technical challenges, as discussed below.
The present 50\% polarization results from muon production in surface muon decay, followed by thermalization in silica aerogel and muonium formation, where hyperfine-induced oscillations depolarize the mixed spin states during the $\sim$100~ns diffusion time. Optical pumping of muonium prior to ionization offers a path to recover and enhance polarization:
a sequence of laser pulses producing Lyman-$\alpha$ excitation and spontaneous
emission cycles can partially select an orientation in the ground
state,
producing a highly polarized ultra-cold muon beam after 355~nm laser ionization. The method requires intense circularly polarized 122~nm pulses, VUV-compatible optics, and sub-microsecond operation compatible with the muon lifetime. Simulations indicate that 70–80\% net polarization may be feasible without modifications to the downstream beamline. A complementary upgrade path is muon acceleration to higher energies: the current linac delivers 212~MeV using RFQ, IH-DTL, DAW-CCL, and DLS structures, but reaching $p_\mu\!\approx\!600$~MeV/$c$ would require a new S-band section possibly combined with a C-band for a required gradient of ~35~MV/m. This is mainly constrained by the limited available space in the accelerator building, which is already tightly packed with linac modules and the storage ring. This high-energy upgrade should be complemented by the increase of the storage-field strength to 6~T. The present superconducting magnet, based on NiTi coils and precision pole shaping, is limited by iron saturation and the requirement of 1ppm field uniformity. Pushing the field toward 6~T would necessitate stronger kickers, larger or differently designed yokes, higher cryogenic capacity, and, if necessary, new coil materials. 
Ongoing studies are evaluating the feasibility of higher-field operation, including field uniformity, mechanical constraints, shielding, and compatibility with injection and beam dynamics, as well as the other potential upgrades.

\section*{Acknowledgements}
I gratefully acknowledge the support of the Leverhulme Trust (grant LIP-2021-014) and the Istituto Nazionale di Fisica Nucleare (Italy).

\end{document}